\documentclass[11pt,english]{article}
\pdfoutput=1
\usepackage[T1]{fontenc}
\usepackage{amssymb}
\usepackage{amsfonts}
\usepackage{amsmath}
\usepackage{cite}
\usepackage{appendix}
\usepackage[unicode=true,
 bookmarks=true,bookmarksnumbered=false,bookmarksopen=false,
 breaklinks=false,pdfborder={0 0 1},backref=false,colorlinks=true]
 {hyperref}
\usepackage{breakurl}
\usepackage[hang,flushmargin]{footmisc}
\usepackage{color}
\usepackage{fullpage}
\usepackage{graphicx}
\usepackage{bm}
\usepackage{tikz,tikz-feynman,float,slashed}

\usepackage{chngcntr}
\counterwithin{equation}{section}

\hypersetup{pdftitle={Title},
 pdfauthor={Authors},
 citecolor=black,linkcolor=black,urlcolor=black}

\newcommand{\email}[1]{\href{mailto:#1}{\nolinkurl{#1}}}

\makeatletter
\def\simgt{\mathrel{\lower2.5pt\vbox{\lineskip=0pt\baselineskip=0pt
           \hbox{$>$}\hbox{$\sim$}}}}
\def\simlt{\mathrel{\lower2.5pt\vbox{\lineskip=0pt\baselineskip=0pt
           \hbox{$<$}\hbox{$\sim$}}}}
\makeatother

\newtheorem{thm}{Theorem}[section]

\newtheorem{example}[thm]{Example}

\newcommand{\Eq}[1]{Eq.~\eqref{#1}}
\newcommand{\Sec}[1]{Sec.~\ref{#1}}
\newcommand{\App}[1]{App.~\ref{#1}}

\DeclareMathOperator{\Tr}{Tr}

\newcommand{\bra}[1]{\langle #1|}
\newcommand{\ket}[1]{|#1\rangle}
\newcommand{\braket}[2]{\langle #1|#2\rangle}
\newcommand{\ketbra}[2]{| #1 \rangle \langle #2 |}

\newcommand{\Hil}{{\cal H}}

\newcommand{\mrm}[1]{\mathrm{#1}}
\newcommand{\dee}{\mathrm{d}}
\newcommand{\ZZ}{\mathbb{Z}}
\newcommand{\GeV}{\mathrm{GeV}}
\newcommand{\MeV}{\mathrm{MeV}}

\usepackage{xcolor}

\usepackage{soul}
\definecolor{purple}{rgb}{0.5,0,0.5}

\makeatletter

\def\lambdabar{\protect\@lambdabar}
\def\@lambdabar{%
\relax
\bgroup
\def\@tempa{\hbox{\raise.73\ht0
\hbox to0pt{\kern.25\wd0\vrule width.5\wd0
height.1pt depth.1pt\hss}\box0}}%
\mathchoice{\setbox0\hbox{$\displaystyle\lambda$}\@tempa}%
{\setbox0\hbox{$\textstyle\lambda$}\@tempa}%
{\setbox0\hbox{$\scriptstyle\lambda$}\@tempa}%
{\setbox0\hbox{$\scriptscriptstyle\lambda$}\@tempa}%
\egroup
}

\makeatother

\begin{document}

\interfootnotelinepenalty=10000
\baselineskip=18pt

\hfill

\thispagestyle{empty}
\begin{center}
{\LARGE \bf
Superselection Rules, Quantum Error Correction, and Quantum Chromodynamics}\\
\bigskip\vspace{1cm}{
{\large Ning Bao${}^{a,b}$, ChunJun Cao${}^{c,d}$, Aidan Chatwin-Davies${}^{e,f}$, Gong Cheng${}^{g}$, Guanyu Zhu${}^{h}$}
} \\[7mm]
 {\it
 ${}^a$Computational Science Initiative, Brookhaven National Lab, Upton, NY  11973, USA\\[1.5mm]
 ${}^b$Department of Physics, Northeastern University, Boston, MA 02115, USA\\[1.5mm]
 ${}^c$Institute for Quantum Information and Matter, Caltech, Pasadena, CA 91125, USA\\[1.5mm]
 ${}^d$Department of Physics, Virginia Tech, Blacksburg, VA 24060, USA\\[1.5mm]
 ${}^e$Department of Physics and Astronomy, University of British Columbia\\[-1mm]
 6224 Agricultural Road, Vancouver, BC, V6T 1Z1, Canada\\[1.5mm]
 ${}^f$Institute for Theoretical Physics, KU Leuven\\[-1mm]
 Celestijnenlaan 200D B-3001 Leuven, Belgium \\[1.5mm]
 ${}^g$Maryland Center for Fundamental Physics, University of Maryland\\[-1mm]
 College Park, MD 20740, USA \\[1.5mm]
 ${}^h$IBM Quantum, T.J. Watson Research Center, Yorktown Heights, NY 10598, USA }

 \let\thefootnote\relax\footnote{\noindent e-mail: \email{ningbao75@gmail.com}, \email{cjcao@vt.edu}, \email{aechatwi@gmail.com}, \email{gongchphy@gmail.com}, \email{guanyu.zhu@ibm.com}} \\
 \bigskip\vspace{0.5cm}{\today}
 \end{center}
\bigskip
\centerline{\large\bf Abstract}
\begin{quote} \small
We investigate the relationship between superselection rules and quantum error correcting codes. We demonstrate that the existence of a superselection rule implies the Knill-Laflamme condition in quantum error correction. As an example, we examine quantum chromodynamics through the lens of quantum error correction, where the proton and neutron states in the model are explored as different superselection sectors that protect logical information.
Finally we comment on topological quantum error correcting codes and supersymmetric quantum field theory within this framework.
\end{quote}

\setcounter{footnote}{0}

\newpage
\tableofcontents
\newpage

\section{Introduction}

Connecting quantum error correction and the AdS/CFT correspondence \cite{maldacena1999large, witten1998anti} has proven to be very fruitful.
The key idea underlying this connection is to interpret the holographic map from the AdS bulk to the CFT boundary as a redundant encoding that has error correcting properties.
While this connection was initially leveraged to resolve the commutator puzzle in AdS/CFT \cite{Almheiri_2015}, it has since led to further advances in the field.
Examples include, but are not limited to, progress in our understanding of entanglement wedge reconstruction \cite{cotler2019entanglement, dong2016reconstruction, faulkner2017bulk}, the development of toy models for holography \cite{Pastawski_2015}, and more realistic refinements of these models \cite{Bao_2019, Cao_2021, Hayden_2016}.
Conversely, these models of holography have inspired new classes of quantum error correcting codes \cite{holosteane,tnc,CL2021,LTNC,Milekhin,largeNcode}.
Going beyond quantum error correction, recently, fault tolerance of the AdS bulk to CFT boundary encoding was investigated and used to relate the confinement/deconfinement phase transition in the CFT to the Hawking/Page phase transition in the AdS gravity theory \cite{Bao_2022}.

First, the success of quantum error correction in AdS/CFT naturally leads to the question of whether techniques from quantum error correction are applicable more broadly in high energy physics.
Beyond the context of holographic conformal field theories, are there further systems or theories that give rise to non-trivial quantum error correcting codes (QECCs), and what insights into these systems and theories do such codes give?
Conversely, one should also expect such connections to inspire new ideas and directions for quantum error correction in and of itself.

Second, a question that was raised in Ref.~\cite{Bao_2022} is whether the confinement/deconfinement phase transition in \emph{any} confining quantum field theory admits an interpretation as a threshold for fault tolerance.
In addition to holographic conformal field theories, this appears to be the case is another special regime of quantum field theory, namely topological quantum field theory, via topological QECCs \cite{aharonov1999faulttolerant, Knill_1998,Kitaev_2003}.
Having a QECC in hand is a prerequisite for discussing fault tolerance, however, and so we are returned to the question of whether other systems---but now, specifically confining field theories---give rise to QECCs.

Demonstrating that any confining quantum field theory admits an interpretation as a QECC is still an extremely general and undoubtedly difficult task, not to mention that it remains a nontrivial step away from proving a threshold theorem for fault tolerance.
Therefore, in this note we will consider a closely related but narrower task, which is to investigate the error correcting properties of theories that have superselection rules.

Given a Hilbert space, a superselection rule (SSR) specifies a partition of the Hilbert space into a direct sum of subspaces called superselection sectors.
The SSR itself consists of the following conditions: no physical operator is allowed to map a state in one sector to a state in another sector, and any physical state cannot consist of a superposition of states from different sectors.
Intuitively, SSRs formalize notions such as how it is impossible to create global superpositions of states with different electric charges, or that there is no such thing as an absolute orientation for angular momentum.
For electric charge (or more generally $U(1)_Y$ hypercharge at high energies) the SSR is thought to be exact, meaning that it is never broken at any scale.
However, a SSR can instead be approximate and only hold in a specific regime of a theory; we will revisit such SSRs in low-energy quantum chromodynamics (QCD) later. 
As any confining quantum field theory necessarily requires a block-diagonal form for its unitary interactions to prevent the direct interaction of non-compatible hadronic sectors at leading order, any such quantum field theory can be shown to respect a SSR.

That a SSR gives rise to a partition of Hilbert space into sectors that cannot be connected by physical operators is interesting from the perspective of quantum information processing.
For instance, SSRs have previously been studied in the context of cryptographic protocols \cite{mayers2002superselection,verstraete2003quantum,Kitaev:2003zj}, where it was hoped that communication in the presence of a SSR could result in stronger cryptographic security. (This hope is unfortunately not borne out, essentially because it is always possible for an adversary to simulate a SSR-violating protocol using SSR-obeying systems \cite{Kitaev:2003zj,Aharonov:1967zza,bartlett2003entanglement}.)

In contrast, SSRs in the context of quantum error correction is a topic that is ripe for investigation.
SSRs immediately give rise to a classical code, in the sense that superselection sectors offer protection against bit flips if codewords are chosen to be representatives of the different sectors.
However, a priori there is no canonical way to protect against phase flips; some further thought is necessary.

For a more concrete model that ties well with our motivation, we examine QCD and related models and how they can be connected to QECCs. As it was shown recently that  $SU(N)$ gauge symmetries at large $N$ can give rise to approximate QECCs \cite{Bao_2022,Milekhin,largeNcode}, QCD is a natural next step in addressing the above motivations because it is a finite $N$ instance of a non-abelian $SU(N)$ gauge theory which exhibits a confinement-deconfinement phase transition. In the low energy regime, it also sustains approximate symmetries, e.g. isospin symmetry, that give rise to approximate SSRs. Therefore, we ask whether one can define a QCD QECC that yields good error correction properties, and whether the confinement-deconfinement transition can correspond to error thresholds like the case in the holographic CFTs. We define a set of QCD codes by leveraging isospin symmetry and, as observed above, we find that these codes are robust against bit-flip errors but susceptible to phase errors. We then explore additional encoding schemes to address phase errors. As expected, this code no longer protects the encoded information above the deconfinement phase transition.

In \Sec{sec:QECC-SSR} we begin by more carefully reviewing the definition of a SSR and relevant prerequisite topics in quantum error correction.
We then clarify how SSRs provide the basic scaffolding for a QECC, as well as the associated limitations, which we illustrate with a simple  example.
Next, in \Sec{sec:QCD}, we examine a richer example of a theory with SSRs, namely low-energy QCD, and we explore its quantum error correcting properties.
In \Sec{sec:discussion} we then conclude with a discussion of further topics, such as topological quantum error correction, supersymmetry, as well as how the present work relates to other literature.

\section{Criteria for error correction from superselection rules}
\label{sec:QECC-SSR}

Let us begin by defining superselection more carefully, following the definitions and conventions of Ref.~\cite{Kitaev:2003zj}.
Let $\Hil$ be a Hilbert space.
A \emph{superselection rule} (SSR) is a decomposition of $\Hil$ into a direct sum of subspaces called \emph{sectors} that are preserved by local operations.
We write
\begin{equation}
    \Hil = \Hil_{q_1} \oplus \Hil_{q_2} \oplus \cdots
\end{equation}
where the label of each sector, $q_i$, is called a \emph{charge}.
The decomposition is determined by specifying a set of operators $\{A_\alpha\}$, called \emph{local operators}, which are such that
\begin{equation}
    \bra{\psi} A_\alpha \ket{\phi} = 0 \qquad \forall ~ ~ \ket{\psi} \in \Hil_{q_i}, ~ \ket{\phi} \in \Hil_{q_j}, ~ q_i \neq q_j.
\end{equation}
Note that the term ``local'' need not have anything to do with real space; at this point, it is just a definition.

This structure is already enough to make contact with error correction via the Knill-Laflamme criterion \cite{knill1997theory}, which we now briefly recall.
Given a physical Hilbert space $\Hil$, and a code subspace $\mathcal{H}_{\rm code}\subset \mathcal{H}$, let $\ket{\bar{i}}\in \mathcal{H}_{\rm code}$ for $ i = 1, ..., n$ be any set of orthonormal vectors, or codewords, such that $\Hil_\mrm{code} = \mrm{span}\{\ket{\bar i}\}_{i=1}^n$.
Given a set of unitary errors $\mathcal{E} = \{E_a\}_{a = 1}^{N}$ that act on states in $\Hil$, the Knill-Laflamme condition states that $\Hil_\mrm{code}$ is a quantum error-correcting code (QECC) that can correct an error drawn from $\mathcal{E}$ if and only if 
\begin{equation} \label{eq:KL}
    \bra{\bar{j}} E_b^\dagger E_a \ket{\bar{i}} = C_{ab} \delta_{ij},
\end{equation}
where $C_{ab}$ is some Hermitian matrix.

Given a SSR, an immediate way to make contact with the Knill-Laflamme condition is to build a code subspace out of (representatives of) superselection sectors, i.e.
\begin{equation}
    \Hil_\mrm{code} = \mrm{span} \{ \ket{q_i} \}_{i \in \Lambda},
\end{equation}
where each $\ket{q_i} \in \Hil_{q_i}$, $\braket{q_i}{q_j} = \delta_{ij}$, and $\Lambda$ is a subset of the charges.
Then, $\Hil_\mrm{code}$ defined in this way can correct errors that are drawn from a subset $\mathcal{E} \subset \{A_\alpha\}$  of the local operators that respect the SSR.
This easily produces the $\delta_{ij}$ part of Eq.~\eqref{eq:KL}.

More precisely, such a code is robust against bit-flip errors; the superselection rule protects the codeword from changing.
However, codewords are still vulnerable to phase-flip errors, since operators such as
\begin{equation} \label{eq:phase-error}
    Z_{q_i} = I - 2 \ketbra{q_i}{q_i}
\end{equation}
are allowed by the SSR and as such are not part of the error model $\mathcal{E}$ defined above.\footnote{A natural question to have at this point would be how superpositions of codewords can be prepared, if superselection rules are so strict as to prevent interaction between codeword sectors. We will return to this point shortly.}
There is no canonical way (in other words, defined by the SSR alone) to protect against such errors.
Nevertheless, for specific phase-flip error models, it is  possible to design codes that offer further protection, which we illustrate with a simple but somewhat artifical construction below.

\subsection{A simple example}
\label{sec:simple-example}

One way in which SSRs can arise is through symmetries.
For instance, let $G$ be a compact group, suppose that $\Hil$ transforms under a unitary representation of $G$, and let the irreducible representations of $G$ label the superselection sectors of $\Hil$.
Following an example from Ref.~\cite{Kitaev:2003zj}, for $G = U(1)$, let the superselection sectors be labelled by the eigenvalues $q \in \mathbb{Z}$ of the charge operator $Q$, with corresponding orthonormal eigenstates $\ket{q}$.
To be even more concrete, $\Hil$ could be the Hilbert space of a particle moving on a unit circle.
In this case, the charge is the momentum of the particle.
Denoting the particle's position on the circle by $\theta$, we can write the momentum eigenstates as
\begin{equation}
    \ket{q} = \frac{1}{\sqrt{2\pi}}\int_0^{2\pi} \dee\theta~e^{iq\theta} \ket{\theta},
\end{equation}
i.e. having a $2\pi$-periodic wavefunction $q(\theta) = e^{iq\theta}/\sqrt{2\pi}$ with respect to which the charge operator has the representation $Q \equiv -i \, d/dq$.
The phase $\theta$ is of course not measurable---there is no absolute notion of position on the circle.
Equivalently, one can also observe that the improper definite phase states
\begin{equation}
    \ket{\theta} = \frac{1}{\sqrt{2\pi}} \sum_{q=-\infty}^{\infty} e^{-iq\theta} \ket{q},
\end{equation}
which satisfy $\braket{\theta}{\theta'} = \delta(\theta-\theta')$, are eigenstates of the charge-nonconserving operator
\begin{equation}
    U_+ = \sum_{q=-\infty}^{\infty} \ket{q+1}\bra{q},
\end{equation}
which does not respect the SSR.
As observed above, if we let each $\ket{q}$ be a codeword, then transitions to different charge states are prevented by the SSR, but nothing stops any given codeword from acquiring an overall phase.

To protect against phase flips, let us consider building a QECC out of the global superselection sectors of several degrees of freedom.
Instead of a single degree of freedom labelled by $q$, suppose that we have two charged degrees of freedom, $A$ and $B$, so that $\Hil = \Hil_A \otimes \Hil_B$.
Now, conservation of charge is the statement that
\begin{equation}
Q_{AB} \equiv Q_A \otimes I_B + I_A \otimes Q_B
\end{equation}
is conserved.
The space $\Hil_q$ of states of definite total charge $q$ is correspondingly
\begin{equation}
\Hil_q = \mrm{span} \{ \ket{\tilde q}_A \ket{q-\tilde q}_B ~ | ~ \tilde q \in \mathbb{Z} \}.
\end{equation}

For each $q \in \mathbb{Z}$, let a codeword be
\begin{equation} \label{eq:naivecode}
\ket{q}_{AB} \equiv \sum_{\tilde q = -\infty}^{\infty} c_{q \tilde q} \, \ket{q-\tilde q}_A \ket{\tilde q}_B,
\end{equation}
where $c_{q\tilde q}\neq 0$ and $\sum_{\tilde q} |c_{q \tilde q}|^2 = 1$.
Consider now some logical state $\ket{\psi}_{AB} = \alpha \ket{q_1}_{AB} + \beta \ket{q_2}_{AB}$.
This state can be written in several ways:
\begin{align}
\ket{\psi}_{AB} &= \alpha \sum_{\tilde q_1} c_{q_1 \tilde q_1} \, \ket{q_1-\tilde q_1}_A \ket{\tilde q_1}_B + \beta \sum_{\tilde q_2} c_{q_2 \tilde q_2} \, \ket{q_2-\tilde q_2}_A \ket{\tilde q_2}_B \label{eq:totalstate} \\
&= \sum_{\tilde q} \left( \alpha c_{q_1 \tilde q} \ket{q_1 - \tilde q}_A + \beta c_{q_2 \tilde q} \ket{q_2 - \tilde q}_A \right) \otimes \ket{\tilde q}_B \label{eq:measureB} \\
&= \sum_{\tilde q} \ket{\tilde q}_A \otimes \left( \alpha c_{q_1 \tilde q} \ket{q_1 + \tilde q}_B + \beta c_{q_2 \tilde q} \ket{q_2 + \tilde q}_B \right).
\end{align}
First, let us suppose that some number of phase flips $Z_q$ gets applied to only one of the degrees of freedom, say $B$.
In that case, if we measure $B$, then $AB$ is projected onto a single term in Eq.~\eqref{eq:measureB}, which can be unitarily rotated to
\begin{equation} \label{eq:recovered}
(\alpha \ket{q_1 - \tilde q}_A + \beta \ket{q_2 - \tilde q}_A)\otimes\ket{\tilde q}_B.
\end{equation}
We thus recover the state $\ket{\psi}_A$ on $A$, at least for this very specific error model in which phase flips are only applied to $B$.

If we did not know whether the phase flips were applied to only $A$ or only $B$, then we could guess.
In the event of a wrong guess (e.g. we measure $B$ but the phase flips happened on $A$), then some of the possible post-measurement and post-rotation states will be of the form
\begin{equation}
(\alpha \ket{q_1 - \tilde q}_A - \beta \ket{q_2 - \tilde q}_A)\otimes\ket{\tilde q}_B,
\end{equation}
i.e. they contain a phase error.
However, provided that only a finite number of phase flips are applied, then the probability that we end with a logical error is vanishing.
This is simply due to the fact that formally the logical states \eqref{eq:naivecode} are a sum over infinite numbers of charge states.
For practical purposes, one would not expect such sums to be possible, and so in practice there would be a finite error rate that is set by the number of charge states that can be prepared in superposition to build a logical state.
In any case, the considerations above illustrate that protecting against phase flips is not generic, but rather depends sensitively on the types of phase flips that can occur.

\subsection{Superpositions of charge states}
\label{sec:charge-superpos}

Thus far we have been writing down superpositions of different charge states rather cavalierly.
A superselection rule, however, by definition prohibits the creation of superpositions of quantum states belonging to different charge sectors.
One way to address this point is to note that, in practice, one can simulate such SSR-violating states if given access to multiple charge-carrying degrees of freedom, similarly to the model for protecting against phase-flip errors above.

As before, let $\Hil$ be a Hilbert space with a SSR that is generated by a compact group $G$.
Given a SSR-violating operator $M$ on $\Hil$ one can write down a charge-invariant \emph{simulation} of $M$ on two copies of the Hilbert space that we here label $\Hil_R$ and $\Hil_S$ \cite{Kitaev:2003zj}:
\begin{equation}
    M^\text{inv} = \sum_{g \in G} (\ketbra{g}{g})_R \otimes (U(g) M U(g)^{-1})_S
\end{equation}
$M^\text{inv}$ is a simulation of $M$ in the sense that for any density matrices $\rho$ and $\rho_R$, it follows that $\Tr(M^\text{inv} \rho_R \otimes \rho) = \Tr(M \rho)$, and $M_1^\text{inv} M_2^\text{inv} = (M_1 M_2)^\text{inv}$ for any operators $M_1$ and $M_2$.\footnote{For proofs of these and further properties, see Ref.~\cite{Kitaev:2003zj}.}
Returning to the simple example with $G = U(1)$, letting $\ket{g} \equiv \ket{\theta}$ and $U(g) \equiv U(\theta) = e^{-i\theta Q}$, we may write
\begin{equation}
    M^\text{inv} = \int_0^{2\pi} \dee \theta ~ (\ketbra{\theta}{\theta})_R \otimes (e^{-i \theta Q} M e^{i \theta Q})_S.
\end{equation}
Then, for a particular superposition of charge states $\ket{\psi} = \sum_q \alpha_q \ket{q}$, we may, for example, prepare a simulation of $\ket{\psi}$ from the $Q_{RS} = 0$ state $\ket{0}_R \ket{0}_S$ by choosing $\bra{q} M \ket{0} = \alpha_q$ and acting with $M^\text{inv}$:
\begin{equation}
    M^\text{inv} \ket{0}_R \ket{0}_S = \sum_q \alpha_q \ket{-q}_R \ket{q}_S
\end{equation}
Should we further wish to incorporate the phase-flip model discussed above, consider making the replacement $S \rightarrow AB$ and replacing each charge state $\ket{q}_S$ with the corresponding codeword $\ket{q}_{AB}$ from \Eq{eq:naivecode}, so that the total simulated state is
\begin{equation} \label{eq:finalRAB}
    \ket{\bar \psi} \equiv \sum_{q,\tilde{q}} \alpha_q c_{q\tilde{q}} \, \ket{-q}_R \otimes \ket{q-\tilde q}_A \otimes \ket{\tilde q}_B.
\end{equation}
By making measurements of $A$ or $B$, we can carry out the same sort of protocol to protect against phase flips as described in Eqs.~\eqref{eq:totalstate}-\eqref{eq:recovered}.
Here, the state $\ket{-q}_R$ simply comes along for the ride once $\tilde q$ is fixed by measuring $A$ or $B$.

The code developed here is a simple  example of how superselection rules and quantum error correction can interact and relies on specific assumptions about the error model.
For instance, an encoded state of the form \eqref{eq:finalRAB} will only retain the bit-flip protection afforded by the SSR as long as $R$, $A$, and $B$ are not allowed to exchange charge once the state has been prepared.
Additionally, the phase-flip protection relies on $R$ and at least one of $A$ or $B$ remaining clean.

As an alternative, we could instead consider building codes out of a SSR that only holds approximately.
For example, suppose that a given SSR is approximate in the sense that it holds true at sufficiently low energies, but can be violated by operators at higher energy scales.
Schematically, one could envision starting with a low-energy state, acting with high-energy operators to bring it above the threshold where the SSR is no longer obeyed, performing a desired logical operation, and then cooling the state back down below the SSR-respecting threshold.

In the following section, we will examine this latter strategy and provide a concrete example of how the correspondence we established earlier between SSRs and the Knill-Laflamme condition for quantum error correction can be instantiated. We will use low energy quantum chromodynamics (QCD) as an illustration, which is a quantum field theory model that approximately follows a SSR. Based on our principle, this implies that the QCD model can be transformed into a quantum error correcting code. 

Subsequently, we will provide a detailed implementation of how to construct the code and demonstrate its non-trivial distance.  It is important to note, however, that building this code for practical purposes would be challenging. Additionally, while we prove the quantum error correction ability based on the Knill-Laflamme condition, it does not inherently provide methods for implementing the recovery process.

\section{QCD code}
\label{sec:QCD}

Let us now consider models assembled from low energy quantum chromodynamics (QCD) below the QCD scale and examine how they can be interpreted as quantum error correcting codes (QECCs).
Like in AdS/CFT, these codes are not meant to be practical QECCs for fault-tolerant quantum computation, but can reveal key structures in the theory that are otherwise opaque.

To begin, we consider only up and down quarks and label them by their two-component spinor $q=\begin{pmatrix}
u \\
d
\end{pmatrix}$.  The Lagrangian is
\begin{equation}\label{eqqcd}
    L_{QCD}=\frac{1}{2}\int \dee^3 x\ \bar{q}(\slashed{\partial}-ig\slashed{A}^{\alpha}T^{\alpha})q-\bar{q}M q,
\end{equation}
where $M=\begin{pmatrix} m_u & 0\\ 0 & m_d\end{pmatrix}$ is the mass matrix. The QCD model has two independent scales: the QCD scale $\Lambda_{QCD}$ and the quark mass scale. If $m_u=m_d$ exactly, the model also exhibits an $SU(2)$ symmetry that rotates the up and down quarks according to the transformation $q\rightarrow Vq$ for $V\in SU(2)$, under which the Lagrangian remains unchanged.
This symmetry ensures that there is no transition between the up and down quarks, thereby leading to the conservation of isospin; however, this symmetry is of course only approximate in reality, as the up and down quark masses are slightly different.

At the Lagrangian level, there exists another approximate symmetry, known as axial $SU(2)$, which applies differently to right- and left-handed quarks when the quark mass is small. However, this symmetry is spontaneously broken by quark condensation in the vacuum, resulting in the emergence of a set of pseudo-Goldstone bosons, the pions \cite{Gell-Mann:1960mvl}. The pion mass is related to the quark masses via \cite{PhysRev.175.2195}
\begin{equation}
    m_{\pi}^2\approx B(m_u+m_d),
\end{equation}
for an energy scale $B \sim 3~\GeV$. The measured mass for the up and down quarks is around $3~\MeV$ and pion mass is around $140~\MeV$ at about one half of $\Lambda_{QCD} \sim 330~\MeV$ \cite{10.1093/ptep/ptaa104}.

In addition to the $SU(2)$ global symmetry, an essential component of the system is the $SU(3)$ gauge symmetry. Quarks carry color charges and transform in the fundamental representation under this symmetry. Each quark field, such as $u_a$ and $d_b$, is associated with color indices, denoted here with subscripts.  The $SU(3)$ gauge symmetry imposes a constraint that all physical states must be gauge singlets.  Notably, this constraint applies to composite particles, including protons and neutrons. They can be precisely defined according to
\begin{equation}
\begin{split}
    &p=\epsilon_{abc}u_au_bd_c\\
    &n=\epsilon_{abc}d_ad_bu_c.
\end{split}
\end{equation}
In these expressions, the color indices are anti-symmetrized using the Levi-Civita symbol $\epsilon_{abc}$ to form charge-neutral objects.

The proton and neutron states belong to different superselection sectors at low energy, meaning that they cannot be transformed into each other by any physical process at energy scales much below $\Lambda_{QCD}$. To be more specific, for any auxiliary state $|\phi\rangle$ with definite energy $E$, the matrix element of the time evolution operator $U(t)$ between proton and neutron states contain a theta function,
\begin{equation}\label{theta}
    \langle\phi'| \langle n|U(t)|p\rangle|\phi\rangle\propto \theta(E-m_{\pi}),
\end{equation}
where $m_{\pi}$ is the mass of pion in the QCD sector, which is around one half of $\Lambda_{QCD}$. Moreover, in the situation in which a proton or neutron is coupled to a thermal bath, and the auxiliary state is in equilibrium at some temperature $T$, the transition amplitude can be estimated as

\begin{equation}\label{eqsuperselection}
   \langle\phi'_T| \langle n|U(t)|p\rangle|\phi_T\rangle\sim O(e^{-\frac{m_{\pi}}{T}}).
\end{equation}
It is suppressed exponentially by a factor $\sim e^{-\frac{\Lambda_{QCD}}{2T}}$. In other words, the probability of a transition between proton and neutron states at temperature  much lower than $\Lambda_{QCD}$ is greatly reduced.

Despite  this approximate superselection rule, it is still possible to make a superposition of proton and neutron states at high enough energies, around the QCD scale\footnote{There are other SSRs in the Standard Model, e.g. the one given by U(1) electric charge conservation, which would prohibit a $p$-$n$ superposition. We will discuss this point in \Sec{SM}.} ($E\sim\Lambda_{QCD}$). This is because at these energies, the quarks inside the proton and neutron can exchange roles. In other words, at high enough energies, the distinction between protons and neutrons becomes fuzzy, and it is no longer possible to uniquely identify which type of baryon is present.

Consider the state $|\psi\rangle=c_1|p\rangle+c_2|n\rangle$. We can encode one qubit of information into this state by assigning logical values 0 and 1 to the proton and neutron states, respectively. 
If we couple this state to a  thermal reservoir at energy much lower than the QCD scale, bit-flip errors are unlikely to occur due to the suppression factor in Eq.~\eqref{eqsuperselection}.
This will be the basis for interpreting the model as a QECC.

However, phase-flip errors can still occur and depend on the specific dynamics of the model. For example, if the particle is coupled to an external electromagnetic field, it can experience a phase shift that causes the encoded information to become corrupted. Therefore, it is important to consider the specific physical environment in which the qubit is placed and to design error correcting codes that can protect against phase-flip errors, much as we saw in \Sec{sec:simple-example}. 

In following subsections, we will discuss two distinct scenarios.  In the first case,  we examine a pure QCD model which is coupled to isospin singlet particles.  Although the interactions may not be physical, this toy model effectively demonstrates the effect of superselection across variable energy scales.   We then move on to the second case, in which we aim to incorporate additional Standard Model interactions, thereby creating a more realistic model. We will show that the practical considerations, although complex, do not impact the primary conclusion drawn from analyzing the simplified model.

\subsection{Toy Model}
\subsubsection*{Approximate superselection rule}
In  Eq.~\eqref{eqqcd} we have the Lagrangian $L_{QCD}$ for the pure QCD sector, and here let us take $m_u=m_d$. To couple it with other fields, we introduce isospin singlet particles $\phi_i$ that have masses $m_i\ll \Lambda_{QCD}$. The full toy-model Lagrangian that we will consider is
\begin{equation}
    L=L_{QCD}+\frac1 2\sum_{i=1,2}[(\partial \phi_i)^2-m_i^2\phi_i^2]+ 
     \lambda_1 \bar{p}p\phi_1+\lambda_2\bar{n}n \phi_2.
\end{equation}
The $\phi_i$ particles are the only constituents of low-energy errors at scales $E\ll \Lambda_{QCD}$. Since they are isospin singlets, they do not induce any transitions between the proton and neutron states. 

As the energy increases, more particles within the QCD sector become excited. Among these excitations, the lightest ones possessing non-trivial isospin are the pions. Pions are composite particles composed of quarks and anti-quarks, and their masses $m_{\pi}$ are approximately one-half of $\Lambda_{\text{QCD}}$. Due to their non-trivial isospin, pions can induce transitions between protons and neutrons. The interaction between pions and baryons can be derived using the effective theory method of QCD \cite{ecker1994pionnucleon,GASSER1984142,SCHWINGER1957407}. In \App{pion}, we present the action and utilize it to calculate the transition amplitude between baryons, providing justification for Eq.~\eqref{theta}.

When the model is coupled to an environment at a finite temperature $T$, the excitation of pions occurs at a rate that is exponentially suppressed by $e^{-\frac{m_{\pi}}{T}}$. This suppression factor arises due to the relatively high mass of the pions compared to the typical temperatures considered in physical systems. Consequently, the probability of pion excitations is significantly reduced.

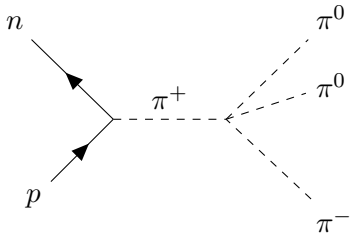
\begin{figure}[H]
\centering
\begin{tikzpicture}
\begin{feynman}
    \vertex (a) {\(p\)};
    \vertex [above right= of a](b);
    \vertex [above left= of b](c){\(n\)};
    \vertex [right = of b] (d);
    \vertex [below right = of d](e){\(\pi^-\)};
    \vertex [above right = of d](f){\(\pi^0\)}; 
    \vertex [below = 0.9cm of f](g){\(\pi^0\)};
    \diagram*{
    (a)--[fermion](b)--[fermion](c),
    (b)--[scalar,edge label=\(\pi^+\)](d),
    (e)--[scalar](d)--[scalar](f),
    (d)--[scalar](g),
    };
\end{feynman}
\end{tikzpicture}
\caption{Flipping error induced by pion scattering. }\label{fig1}
\end{figure}

When pions are thermally excited, they can interact with protons and neutrons and induce bit-flip errors, as depicted in Fig~\ref{fig1}. This justifies the right hand side of Eq.~\eqref{eqsuperselection}: the proton-neutron superselection rule is approximately valid for low energy $E\ll\Lambda_{QCD}$, but is violated for energy scales comparable or larger than the QCD scale. 

\subsubsection*{Phase errors}
The naive superposition of $|p\rangle$ and $|n\rangle$ can be used to encode one qubit of information. However, when coupled to an environment, the qubit can be subjected to phase errors. It is therefore necessary to consider more complicated codewords in order to interpret the model as an instance of quantum error correction. 
 
 In general, the errors acting on codewords arise from the unitary time evolution of the combined system of the code states and the ambient modes. Mathematically, we denote a codeword as $\rho_{\psi}$ and the environment modes collectively as $|\phi\rangle$. By inserting specific ambient modes $|\phi\rangle$ we induce corresponding errors $\{E^{\phi}_k\}$.  The general error channel is then
\begin{equation}
    \sum_k E^{\phi}_k\rho_{\psi}E^{\phi \dagger}_k=\Tr_{\phi}( U(t)\rho_{\psi}\otimes|\phi\rangle\langle\phi|U^{\dagger}(t) ).
\end{equation}
 We aim to find the code states, labeled by $|\tilde i\rangle$ and $|\tilde j\rangle$, that satisfy the  Knill-Laflamme condition,\footnote{It remains conjectural that the Knill-Laflamme condition works in the continuum limit (in which lattice spacing goes to zero); however, it is widely expected to not fail. In particular, the work on holographic QECCs \cite{Almheiri_2015} requires it to hold. For the purposes of this work, we will assume the continuum generalization to be correct. Aspects of this question are discussed in \cite{Faist_2020}.} 
 
 \begin{equation}\label{eqKLQCD}
 \langle\tilde{j}|E^{\phi'\dagger}_{k'}E^{\phi}_k|\tilde{i}\rangle=C_{kk'}(\phi,\phi')\delta_{ij}.
 \end{equation}

 As an example, consider the case when  $|\phi\rangle$ is a single particle state and the evolution time is long enough that the error channel can be treated as a scattering process,
 \begin{equation}
 \begin{split}
     &\lim_{t\rightarrow\infty}U(t)|\tilde{i},0\rangle|s,k\rangle=\sum_{k'}\mathcal{A}^{s,\tilde{i}}_{k, k'}|\tilde{i},k'\rangle|s,k-k'\rangle.
\end{split}
 \end{equation}
 We denote the environmental states by $|s,k\rangle$, where $s$ is the particle species and $k$ is its 3-momentum. In this toy model, $s$ can be either $\phi_1$ or $\phi_2$.   The logical state $|\tilde{i}\rangle$ initially has zero-momentum, which we denote by $|\tilde{i},0\rangle$. From the equation above, the error operator is explicitly related to the scattering amplitude via 
 \begin{equation}
 \begin{split}
 &E^{|s,k\rangle}_{k'}|\tilde{i},0\rangle=\mathcal{A}^{s,\tilde{i}}_{k, k'}|\tilde{i},k'\rangle.
 \end{split}
 \end{equation}

 If we use the naive superposition of proton and neutron as logical states, then the superselection property in Eq.~\eqref{eqsuperselection} already gives us the factor $\delta_{ij}$ in the Knill-Laflamme condition Eq.~\eqref{eqKLQCD}. However, the prefactor $C(\phi,\phi')$ can depend on the logical state.  To see this, we simply evaluate the Knill-Laflamme matrix element and write it as the scattering amplitude, 
\begin{equation}
    \langle\tilde{i}|E^{|s,k\rangle \dagger}_{k'}E^{|s,k\rangle}_{k'}|\tilde{i}\rangle=|\mathcal{A}^{s,\tilde{i}}_{k, k'}|^2
\end{equation}

Let us first examine the errors induced by the isospin singlet particles $\phi_1$ and $\phi_2$. Focusing on the case when $k'=0$, one easily obtains the amplitudes 
 
 \begin{equation}
 \begin{split}
     \mathcal{A}^{\phi_1,p}_{k,0}=1-\frac{\lambda_1^2}{2}, \qquad  \mathcal{A}^{\phi_1,n}_{k,0}=1\\
     \mathcal{A}^{\phi_2,p}_{k,0}=1, \qquad \mathcal{A}^{\phi_2,n}_{k,0}=1-\frac{\lambda_2^2}{2}. 
\end{split}
 \end{equation}
Since the amplitudes for protons and neutrons are different, i.e., $\mathcal{A}^{s,p}_{k,k'}\neq \mathcal{A}^{s,n}_{k,k'}$, this scattering process induces a phase error. In fact, we can explicitly see this by expanding the state after the scattering process,

 \begin{equation}
 \begin{split}
     U(c_1|p,0\rangle+c_2|n,0\rangle)|s,k\rangle=&\sum_{k'}(c_1\mathcal{A}_{k,k'}^{s,p}|p,k'\rangle+c_2\mathcal{A}_{k,k'}^{s,n}|n,k'\rangle)|s,k-k'\rangle\\
     =&\sum_{k'}[(\alpha_1(k') I+\alpha_2(k') Z)(c_1|p,k'\rangle+c_2|n,k'\rangle)]\otimes|s,k-k'\rangle.
 \end{split}
 \end{equation}
The state is a superposition of  different branches, each labeled by a particular momentum $k'$. For each $k'$, the error is induced by an error operator $\alpha_1(k')I+\alpha_2(k')Z$, with $\alpha_1$ and $\alpha_2$ given by 
\begin{equation}
\begin{split}
   & \alpha_1(k')=\frac{\mathcal{A}_{k,k'}^{s,p}+\mathcal{A}_{k,k'}^{s,p}}{2},\\
   & \alpha_2(k')=\frac{\mathcal{A}_{k,k'}^{s,p}-\mathcal{A}_{k,k'}^{s,p}}{2}.
\end{split}
\end{equation}

In contrast to the approach in \Sec{sec:simple-example}, here let us correct the phase error by concatenating with a phase-flip (repetition) code. Define 
\begin{equation}
\begin{split}
   & |+,k\rangle:=\frac{|p,k\rangle+|n,k\rangle}{\sqrt{2}}\\
   & |-,k\rangle:=\frac{|p,k\rangle-|n,k\rangle}{\sqrt{2}}
\end{split}
\end{equation}
and encode the logical information in the state of several entangled physical particles, 
\begin{equation}\label{eqrepetition}
  |\tilde{\psi}\rangle=  c_{+}|+,0\rangle|+,0\rangle\cdots |+,0\rangle+c_{-}|-,0\rangle|-,0\rangle\cdots|-,0\rangle.
\end{equation}
Again the errors are induced by scattering between some particle in the environment with this group of particles. We assume that the interaction happens locally and only affects one physical particle each time. For example, let us assume that only the first particle is affected. The error then maps the logical state to

\begin{equation}
\begin{split}
    U|\tilde{\psi}\rangle|s,k\rangle = \sum_{k'}\left[\alpha_1(k')\left(c_{+}|+,k'\rangle\otimes|++\cdots+,0\rangle+c_{-}|-,k'\rangle\otimes|--\cdots-,0\rangle\right)\right. \\ + \left. \alpha_2(k')\left(c_+|-,k'\rangle\otimes|++\cdots+,0\rangle+c_-|+,k'\rangle\otimes|--\cdots-,0\rangle\right)\right]\otimes|s,k-k'\rangle.
\end{split}
\end{equation}

To recover the logical information from this scattered state, we need to make some assumptions. First, assume that we could  measure just the momentum  without perturbing the isospin of particles. This projection would pick one state with definite momentum $k'$ out of the sum in the above equation. Second, assume that we could separate the auxiliary particle $s$ from the other physical particles after the projection (this assumption might not be realistic), so we make sure that the subsequent recovery operations only apply to the physical particles. 

After projecting to a definite momentum, apply a boost to the scattered particle based on the measured momentum value to bring it back to the same momentum as the other physical particles. After this operation, the state becomes

\begin{equation}
\begin{split}
   |\psi\rangle \rightarrow \alpha_1(k')(c_{+}|+++\cdots+,0\rangle+c_{-}|---\cdots-,0\rangle)\\
    +\alpha_2(k')(c_+|-++\cdots+,0\rangle+c_-|+--\cdots-,0\rangle)
 \end{split}
\end{equation}
The error can be detected by measuring the syndrome $\{X_1X_2, X_2X_3\}$ acting on the first three qubits. For more general phase-flip errors, we can apply the standard recovery channel of the phase-flip code to recover the logical information.

\subsection{Generalization with Standard Model interactions }\label{SM}

It is possible to generalize the toy model presented above by including more realistic interactions from the Standard Model, where we have both weak interactions described by the $SU(2)$ gauge theory and electromagnetic interactions described by the $U(1)$ theory. However, there are several complications that arise from including these interactions which affect the model's interpretation as an instance of quantum error correction.

First, the weak interaction induces transitions between the proton and neutron states through interactions with particles in the lepton sector. Specifically, in the neutron spontaneous decay channel, a neutron can decay into a proton, an electron, and an antineutrino through the weak interaction. This can cause an error in the encoding of the qubit, as the neutron state has been converted to a proton state. Similarly, in the electron capture channel of the proton, a proton can capture an electron and be converted to a neutron and a neutrino through the weak interaction, causing a bit-flip error in the encoding. Since leptons have small masses and can be easily created from the environment, these additional sources of errors must be analyzed carefully. 

Second, the $U(1)$ charges of the proton and neutron states are different, and therefore it is impossible to make a direct superposition of these states due to the superselection rule imposed by $U(1)$ charge conservation. This means that a more complex encoding scheme is needed in order to interpret qubits as being encoded in a combined proton-neutron state. 

Despite these complications, the conceptual feasibility of encoding and manipulating qubits in baryon superposition states remains unchanged, based on the validity of the approximate SSR. We will discuss this in greater detail in the following subsections. 

\subsubsection*{Transitions induced by weak interactions}

Protons and neutrons can be converted into one another by interacting with $W$ and $Z$ bosons, whose mass scale is $m_W,m_Z\sim 90~\GeV \gg \Lambda_{QCD}$. In reality, the mass of a neutron is larger than that of a proton; a single neutron spontaneously decays into a proton, an electron and a neutrino, whereas a single proton is stable. However, the neutron decay rate is very small, occurring on a time scale of about 15 mins. Therefore, this effect is negligible for the purpose of preserving encoded quantum information as long as one is satisfied with coherence times that are much shorter than this time scale.

The channel for a proton converting to a neutron is through an electron capture process, shown in the following diagram:

\begin{figure}[H]
    \centering
    \begin{tikzpicture}
    \begin{feynman}
        \vertex (a) {\(p\)} ;
        \vertex [above right= of a](b) ;
        \vertex [above left=of b](c) {\(n\)};
        \vertex [right= of b] (d);
        \vertex [below right= of d] (e){\(e^-\)};
        \vertex [above right= of d] (f){\(\nu_e\)};
        \diagram*{
        (a)--[fermion](b),
        (b)--[fermion](c),
        (b)--[photon, edge label=\(W^+\)](d),
        (e)--[fermion](d),
        (d)--[fermion](f)
        };
    \end{feynman}
    \end{tikzpicture}
    \caption{Bit-flip error induced by lepton scattering mediated by a $W$ boson.}
    \label{fig2}
\end{figure}
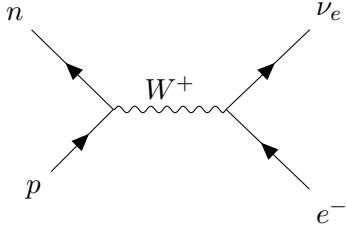

\noindent This process is suppressed by a factor of  $(\frac{E}{m_W})^2$. When $E< 0.05 \, \Lambda_{QCD}$, it follows that $(\frac{E}{m_W})^2>e^{-\frac{\Lambda_{QCD}}{E}}$. In other words, this is the regime in which leptons that carry non-trivial isospin would be the dominant bit-flip error. In contrast, for  $E>0.05 \, \Lambda_{QCD}$, pions are the dominant source of error. 

In summary, the states of protons and neutrons indeed constitute approximate superselection sectors at low energy, even when considering further Standard Model interactions. The suppression factor located on the right side of Eq.~\eqref{eqsuperselection}, however, is modified to $\text{max}(e^{-\frac{\Lambda_{QCD}}{E}}, \frac{E^2}{m_{W}^2})$.  

\subsubsection*{$U(1)$ superselection rule}

Since proton and neutron states carry different $U(1)$ charges, $U(1)$ symmetry imposes a SSR that prevents the superposition of protons and neutrons. The $U(1)$ SSR is still present even at extremely high energy $E\gg \Lambda_{weak}$, when the $SU(2)_L\times U(1)_Y$ is not spontaneously broken by the Higgs mechanism. This forbids the superposition of up and down quarks, because they carry different $U(1)_Y$ hyper-charges, $Q_{u_R}=\frac{4}{3}$, and $Q_{d_R}=-\frac{2}{3}$.

One possible way to address the $U(1)$ SSR is to consider SSR-conforming superpositions of more than one degree of freedom (cf. \Sec{sec:charge-superpos}). In other words, consider superpositions of states in the form of 
\begin{equation}
    |\psi\rangle =\alpha |p\rangle|e^{-}\rangle+\beta|n\rangle|\nu\rangle
\end{equation}
that have both definite electric charge and lepton number. In principle, one could even envision creating a magnetic trap to constrain the proton and electron pair within a localized region. The neutrino has negligible interactions with the environment, so we can safely assume that no error could act on the neutrino state.

\subsubsection*{Bit-flip errors and phase errors}

As we mentioned before, the bit-flip errors in this model are induced by  both the meson interactions (as shown in Fig.~\ref{fig1}) and the electron capture process (as in Fig.~\ref{fig2}), which are suppressed either by the QCD scale or the mass of $W$ and $Z$ bosons. The dominant contribution to phase errors comes from the electromagnetic interaction.  Since the proton and neutron states carry different $U(1)$ charges, coupling to the electromagnetic field can easily induce a relative phase change on $|p\rangle$ and $|n\rangle$, 

\begin{equation}
    |\psi\rangle\rightarrow \alpha e^{-ie \theta}|p\rangle|e^-\rangle+\beta|n\rangle|\nu\rangle
\end{equation}

Nevertheless, a repetition code as in Eq.~\eqref{eqrepetition} would still succeed in protecting the logical information.

\subsection{Distance of the QCD code}

The distance of a quantum error-correcting code (QECC) is typically defined as the minimum number of physical qubits that need to be changed by an error to convert one valid codeword to another. This definition is natural in conventional QECCs based on qubits because the dominant error is local and well-approximated by the depolarizing channel. Both the encoding and the error models are quite different in the case of the QCD code examined here---a single baryon is used to encode information, and the codewords correspond to proton and neutron states.  A pertinent question is how the code distance should be defined in this scenario. 

In a general QECC model, it is commonly assumed that each single-qubit error occurs with equal probability $p$, and is independent of the qubit's polarization. Therefore, an error operator that alters the codeword would occur with a probability no greater than $O(p^L)$, where $L$ is the code distance. 

For the QCD code, in analogy, we can consider the probability of an error occurring with an energy $E$. At a given temperature $T$, this error occurs with thermal probability $\sim e^{-\frac{E}{T}}$. This can be rewritten as $(e^{-\frac{\epsilon}{T}})^{\frac{E}{\epsilon}}$, where $\epsilon$ is the energy scale of a typical process at that temperature. Hence, we can tentatively identify $e^{-\frac{\epsilon}{T}}$ as the probability of a ``single-qubit'' error, denoted as $p$, and the energy measured in units of $\epsilon$ as the size of the error operator. In this framework, we can define the effective distance of the QCD code as $\frac{\Lambda_{QCD}}{\epsilon}$, with $\Lambda_{QCD}$ being the QCD scale, and $\epsilon$ being the scalar particle mass in the toy model and lepton mass in the standard model.

\section{Discussion}
\label{sec:discussion}

Before we make some general comments regarding superselection and quantum error correction, let us first discuss further instances of the relationship between the two.

\subsection{Comparing the QCD code with the $SU(3)$ toric code}

The toric code can be understood as a $\mathbb{Z}_2$ lattice gauge theory \cite{PhysRevD.19.3682}. This theory can be generalized to include a gauge group of $SU(3)$, or more generally, $SU(N)$, as noted in \cite{PhysRevA.105.052423, PhysRevB.79.045316}. This generalization preserves many intriguing properties of the $\mathbb{Z}_2$ toric code, including the existence of topological sectors and the indistinguishability of these sectors when subjected to local operators. As such, like the $\mathbb{Z}_2$ toric code, these generalized versions can also be utilized as QECCs.

In comparison to the QCD code that we constructed earlier, the $SU(3)$ toric code exhibits a clear advantage: it inherently functions as a quantum code. The QCD code functions as a classical code that only protects against bit-flip errors. 

We underscore the fact that the QCD code and the $SU(3)$ toric code correspond to distinct regimes of the $SU(3)$ gauge theory with suitably chosen coupling constants. Both codes can protect information coherently up to a certain temperature. However, the QCD code possesses a much higher temperature threshold, attributable to the significant binding energy of gluons that confine quarks within hadrons. Conversely, the $SU(3)$ toric code model exists in a fully deconfined regime, with a substantially smaller energy gap for excitations.

\subsection{Topological quantum error correcting codes and topological quantum field theory}
Another class of quantum error correcting codes that exhibits a relationship to superselection rules are topological codes.   The code space $\mathcal{H}_\text{code}$ of a topological code is the Hilbert space of a corresponding topological quantum field theory, or equivalently a discrete gauge theory with a discrete gauge group $G$.

The simplest topological code, the toric code,  corresponds to the case of $G=\ZZ_2$. 
Consider more generally the $\ZZ_N$ gauge theory, which can be described with the Chern-Simons action:
\begin{equation}\label{eq:CS}
S_\text{CS} = \int  \frac{N}{2\pi} b \wedge da.
\end{equation}
Here, $a$ and $b$ are 1-form compact $U(1)$ gauge fields describing the electric and magnetic degrees of freedom.  By setting $N=2$ we get back to the $\ZZ_2$ gauge theory and toric code.   There are four superselection sectors in the ground-state subspace, or equivalently code space $\mathcal{H}_\text{code}$ in the toric code, i.e., 
$\mathcal{H}_\text{code}=\text{span}\{\ket{\mathbb{I}}_x, \ket{e}_x, \ket{m}_x, \ket{em}_x \}$, where the basis states $\ket{\mathbf{a}}_x$ are the eigenstates of the Wilson loop operator $W^{\mathbf{a}}_x$ corresponding to charge $\mathbf{a}$ along the $x$-cycle of the torus (the other is cycle is denoted by $y$).
The superselection rule states that for any local operator $A$ with support less than the code distance $d$, one has
\begin{equation}
    _x\bra{\mathbf{a}} A \ket{\mathbf{a}'}_x = 0 \qquad \forall ~ \ket{\mathbf{a}}_x, \ket{\mathbf{a'}}_x \in \mathcal{H}_\text{code},  ~ \mathbf{a} \neq \mathbf{a}',
\end{equation}
which is consistent with the Knill-Laflamme condition.

\subsection{Supersymmetric quantum field theory}
Some work relating quantum error correction to supersymmetric quantum field theory has already been done in \cite{Harvey_2020}. The work in this article is complementary to this previously existing work, and perhaps points to how it can be generalized.

Specifically, supersymmetry generates its own superselection rules governing its dynamics dependent on the number of conserved supercharges \cite{Tachikawa_2015, wess1992supersymmetry}. These superselection rules are broken by breaking supersymmetry from a higher number of conserved supercharges down to a lower number, e.g. from $\mathcal{N}=4$ supersymmetry down to $\mathcal{N}=2$. Per the argument presented in this work, this implies that the supersymmetry-derived superselection rules also admit a quantum error correction interpretation. Indeed, it is well-known that the dynamics of supersymmetric field theories are increasingly restricted by the number of conserved supercharges. This corresponds quite well  to protected logical qubits with a relatively restrictive set of allowed logical gates at higher $\mathcal{N}$ and less well protected logical qubits with a more permissive set of allowed logical operations at smaller or zero $\mathcal{N}$. It is of clear interest to further study the potential ramifications of the quantum error correction picture on the field of supersymmetry as a whole, in particular in the context of supersymmetry breaking and constraints on dynamical processes.

\subsection{Scale-dependent error correction and code concatenation}

In defining low energy subspaces of quantum field theories as the code subspaces, the natural error model associated with the code corresponds to thermal errors, where the dominant errors come from interactions with particles at some energy scale $\mu\sim T$. For instance, we saw that below QCD scale, the possible errors modes were dominated by interactions with the isospin singlet $\phi$ in the toy model or decays mediated by the electroweak interaction in the Standard Model. However, for energies near the QCD scale, they were dominated by the exchange of mesons, such as pions at the 100 MeV scale. Then for $\mu>\Lambda_{QCD}$ we would expect the interactions, and hence primary error modes, to be  dominated by quark-gluon interactions in the asymptotically free regime. In general, we can understand this in terms of the unbroken symmetries at different scales that gave rise to different types of permissible processes. This constitutes an error channel $\mathcal{E}(\mu)$ which now also depends on the energy $\mu$, or equivalently, a length scale $\ell\sim 1/\mu$. In this case, the error model is not just one where the ``physical degrees of freedom'' simply become noisier like in the usual Pauli depolarizing channel, but the type of error that is turned on is also different at different length or energy scales. 

Although the above field theory code or error model has little bearing on practical error correction, it does motivate the possibility that the nature of errors may be length or energy scale dependent in a physical system. For such a system, it is more beneficial to tailor a code that is designed to correct such scale-dependent errors. For instance, one can consider a concatenated quantum code where the block code used at different levels of concatenation is best protected against errors at that scale. This is similar to the observation of RG as a QECC \cite{Furuya_2022,Furuya1_2022}, wherein each RG or coarse-graining step maps to an encoding isometry. Then, irrelevant operators in the UV correspond to correctable errors that do not harm the logical information encoded in the IR.

For concreteness, one could construct a Pauli-based error model as a toy example. Suppose that there are $n$ physical qubits arranged in a 1-dimensional lattice subjected to short-range $Z$ errors and long range $X$ errors such that the ``short-ranged'' errors are dominated by single qubit phase errors $Z$ with some probability $p_Z$ while the longer range errors are correlated $X^{\otimes l}$ of $l$ contiguous bit flips of probability $p_X$. To tailor a code for such an error, one could construct a concatenated code that is a gauge fixed Bacon-Shor code. In the UV level, one could use repetition codes that protect against phase-flip errors and in the IR layer use codes that best protects against bit-flip errors.

In this work, we used code concatenation to convert the current QCD (outer) codes into a quantum code by further encoding the hadronic degrees of freedom with another classical (inner) code that protects against phase errors. The concatenation with the artificial repetition code may not have a natural interpretation in RG. However, one may also leverage other symmetries and associated (S)SRs at a deeper IR scale to identify more natural inner codes that protect against phase flips. For instance, one may be able to identify states of more complicated atomic nuclei as codewords with such desired properties. 

\subsection{General comments and future directions}
There are several general comments worth making regarding the connection between superselection and quantum error correction. First, we can consider the situation in which two different approximate superselection sectors corresponding to independent conservation of two different quantities exist. This would correspond to a concatenated quantum error correcting code, where one can fix the code subspace corresponding to a specific value of the first quantity, and then fix the code subspace corresponding to the second. Such fixing in this case is commutative, something which is not always the case in concatenated quantum codes. It is worth considering in the future if there is a way to use superselection sectors to form non-commutative quantum error correcting codes; this may take the form of conditional superselection rules, e.g. rules that only hold for a given second quantity if the first quantity obeys some constraint, as in points of enhanced symmetry.

In particular, we could specialize to cases in which two approximate superselection rules break down at different energy scales.
This would correspond to a situation in which the quantum error correcting properties of the code corresponding to the lower breakdown energy stops correcting nontrivially first, while the code corresponding to the higher energy remains functional. By specifying the scales in question to be those associated with, e.g., breakings of specific supersymmetries or with the scales appearing in the Standard Model construction described above, one can fold both of these instances of multiscale quantum error correction into this framework.

A specific way that the connection between superselection and quantum error correction could help is in approaching the mass gap problem. Here, the challenge is to study the relationship between the confinement/deconfinement transition and the mass gap of QCD. Based on the QECC picture, certain small sets of errors cannot take a state outside of a superselection sector by breaking the approximately conserved superselection rule, but a sufficiently large number of errors can. Supposing that the mass gap was given by the minimal number of such errors needed to break the approximate SSR from isospin symmetries, then the problem of finding the mass gap can be related to the problem of finding the distance of the QECC. A more rigorous connection between the problems will allow us to tackle the mass gap problem using technologies for analyzing code distances. Interestingly, the problem of finding code distance is also NP-hard\cite{mindist}. Therefore, even if one cannot precisely compute the distance (and by extension, the mass gap), we could in principle establish a complexity theoretic constraint on how hard the mass gap problem can be.

Finally, it is worth mentioning that while we have established a relation between superselection rules and quantum error correction, a full connection between superselection rules and fault tolerance remains elusive. Specifically, we have not established what the criteria for fault tolerance are in the language of superselection, nor what the specific threshold is. A naive attempt at this would be to set the threshold at the scale at which approximate superselection breaks down, but this would need to be examined carefully because the breakdown of superselection is not necessarily equivalent to the breakdown of fault tolerance.

\begin{center} 
{\bf Acknowledgments}
\end{center}
\noindent 
We thank Philippe Faist, Layla Hormozi, Charles Marteau, Brian Swingle, and Christopher White for helpful discussions during the preparation of this manuscript.
ACD acknowledges the support of the Natural Sciences and Engineering Research Council of Canada (NSERC), [funding reference number PDF-545750-2020] and was supported for a portion of this work as a postdoctoral fellow (Fundamental Research) of the National Research Foundation -- Flanders (FWO), Belgium.
N.B. is supported by the Computational Science Initiative at Brookhaven National Laboratory, Northeastern University, and by the U.S. Department of Energy QuantISED Quantum Telescope award. C.C. acknowledges  the Air Force Office of Scientific Research (FA9550-19-1-0360), and the National Science Foundation (PHY-1733907). The Institute for Quantum Information and Matter is an NSF Physics Frontiers Center. G.Z. acknowledges the support by the U.S. Department of Energy, Office of Science,
National Quantum Information Science Research Centers, Co-design Center for Quantum Advantage (C2QA)
under contract number DE-SC0012704.
G.C. acknowledges support from the U.S. Department of Energy, Office of Science, Office of Advanced Scientific Computing Research, Accelerated Research for
Quantum Computing program “FAR-QC”.

\bibliographystyle{utphys-modified}
\bibliography{refs.bib}

\providecommand{\href}[2]{#2}\begingroup\raggedright\begin{thebibliography}{10}

\bibitem{maldacena1999large}
J.~M. Maldacena, ``{The Large N limit of superconformal field theories and
  supergravity},'' \href{http://dx.doi.org/10.4310/ATMP.1998.v2.n2.a1}{{\em
  Adv. Theor. Math. Phys.} {\bfseries 2} (1998) 231--252},
  \href{http://arxiv.org/abs/hep-th/9711200}{{\ttfamily arXiv:hep-th/9711200}}.

\bibitem{witten1998anti}
E.~Witten, ``{Anti-de Sitter space and holography},''
  \href{http://dx.doi.org/10.4310/ATMP.1998.v2.n2.a2}{{\em Adv. Theor. Math.
  Phys.} {\bfseries 2} (1998) 253--291},
  \href{http://arxiv.org/abs/hep-th/9802150}{{\ttfamily arXiv:hep-th/9802150}}.

\bibitem{Almheiri_2015}
A.~Almheiri, X.~Dong, and D.~Harlow, ``{Bulk Locality and Quantum Error
  Correction in AdS/CFT},''
  \href{http://dx.doi.org/10.1007/JHEP04(2015)163}{{\em JHEP} {\bfseries 04}
  (2015) 163}, \href{http://arxiv.org/abs/1411.7041}{{\ttfamily arXiv:1411.7041
  [hep-th]}}.

\bibitem{cotler2019entanglement}
J.~Cotler, P.~Hayden, G.~Penington, G.~Salton, B.~Swingle, and M.~Walter,
  ``{Entanglement Wedge Reconstruction via Universal Recovery Channels},''
  \href{http://dx.doi.org/10.1103/PhysRevX.9.031011}{{\em Phys. Rev. X}
  {\bfseries 9} no.~3, (2019) 031011},
  \href{http://arxiv.org/abs/1704.05839}{{\ttfamily arXiv:1704.05839
  [hep-th]}}.

\bibitem{dong2016reconstruction}
X.~Dong, D.~Harlow, and A.~C. Wall, ``{Reconstruction of Bulk Operators within
  the Entanglement Wedge in Gauge-Gravity Duality},''
  \href{http://dx.doi.org/10.1103/PhysRevLett.117.021601}{{\em Phys. Rev.
  Lett.} {\bfseries 117} no.~2, (2016) 021601},
  \href{http://arxiv.org/abs/1601.05416}{{\ttfamily arXiv:1601.05416
  [hep-th]}}.

\bibitem{faulkner2017bulk}
T.~Faulkner and A.~Lewkowycz, ``{Bulk locality from modular flow},''
  \href{http://dx.doi.org/10.1007/JHEP07(2017)151}{{\em JHEP} {\bfseries 07}
  (2017) 151}, \href{http://arxiv.org/abs/1704.05464}{{\ttfamily
  arXiv:1704.05464 [hep-th]}}.

\bibitem{Pastawski_2015}
F.~Pastawski, B.~Yoshida, D.~Harlow, and J.~Preskill, ``{Holographic quantum
  error-correcting codes: Toy models for the bulk/boundary correspondence},''
  \href{http://dx.doi.org/10.1007/JHEP06(2015)149}{{\em JHEP} {\bfseries 06}
  (2015) 149}, \href{http://arxiv.org/abs/1503.06237}{{\ttfamily
  arXiv:1503.06237 [hep-th]}}.

\bibitem{Bao_2019}
N.~Bao, G.~Penington, J.~Sorce, and A.~C. Wall, ``{Beyond Toy Models:
  Distilling Tensor Networks in Full AdS/CFT},''
  \href{http://dx.doi.org/10.1007/JHEP11(2019)069}{{\em JHEP} {\bfseries 11}
  (2019) 069}, \href{http://arxiv.org/abs/1812.01171}{{\ttfamily
  arXiv:1812.01171 [hep-th]}}.

\bibitem{Cao_2021}
C.~Cao and B.~Lackey, ``{Approximate Bacon-Shor Code and Holography},''
  \href{http://dx.doi.org/10.1007/JHEP05(2021)127}{{\em JHEP} {\bfseries 05}
  (2021) 127}, \href{http://arxiv.org/abs/2010.05960}{{\ttfamily
  arXiv:2010.05960 [hep-th]}}.

\bibitem{Hayden_2016}
P.~Hayden, S.~Nezami, X.-L. Qi, N.~Thomas, M.~Walter, and Z.~Yang,
  ``{Holographic duality from random tensor networks},''
  \href{http://dx.doi.org/10.1007/JHEP11(2016)009}{{\em JHEP} {\bfseries 11}
  (2016) 009}, \href{http://arxiv.org/abs/1601.01694}{{\ttfamily
  arXiv:1601.01694 [hep-th]}}.

\bibitem{holosteane}
R.~J. Harris, N.~A. McMahon, G.~K. Brennen, and T.~M. Stace,
  ``{Calderbank-Shor-Steane holographic quantum error-correcting codes},''
  \href{http://dx.doi.org/10.1103/PhysRevA.98.052301}{{\em Phys. Rev. A}
  {\bfseries 98} no.~5, (2018) 052301},
  \href{http://arxiv.org/abs/1806.06472}{{\ttfamily arXiv:1806.06472
  [quant-ph]}}.

\bibitem{tnc}
T.~Farrelly, R.~J. Harris, N.~A. McMahon, and T.~M. Stace, ``{Tensor-Network
  Codes},'' \href{http://dx.doi.org/10.1103/PhysRevLett.127.040507}{{\em Phys.
  Rev. Lett.} {\bfseries 127} no.~4, (2021) 040507},
  \href{http://arxiv.org/abs/2009.10329}{{\ttfamily arXiv:2009.10329
  [quant-ph]}}.

\bibitem{CL2021}
C.~Cao and B.~Lackey, ``{Quantum Lego: Building Quantum Error Correction Codes
  from Tensor Networks},''
  \href{http://dx.doi.org/10.1103/PRXQuantum.3.020332}{{\em PRX Quantum}
  {\bfseries 3} no.~2, (2022) 020332},
  \href{http://arxiv.org/abs/2109.08158}{{\ttfamily arXiv:2109.08158
  [quant-ph]}}.

\bibitem{LTNC}
T.~Farrelly, D.~K. Tuckett, and T.~M. Stace, ``{Local tensor-network codes},''
  \href{http://dx.doi.org/10.1088/1367-2630/ac5e87}{{\em New J. Phys.}
  {\bfseries 24} no.~4, (2022) 043015},
  \href{http://arxiv.org/abs/2109.11996}{{\ttfamily arXiv:2109.11996
  [quant-ph]}}.

\bibitem{Milekhin}
A.~Milekhin, ``{Quantum error correction and large $N$},''
  \href{http://dx.doi.org/10.21468/SciPostPhys.11.5.094}{{\em SciPost Phys.}
  {\bfseries 11} (2021) 094}, \href{http://arxiv.org/abs/2008.12869}{{\ttfamily
  arXiv:2008.12869 [hep-th]}}.

\bibitem{largeNcode}
C.~Cao, G.~Cheng, and B.~Swingle, ``{Large $N$ Matrix Quantum Mechanics as a
  Quantum Memory},'' \href{http://arxiv.org/abs/2211.08448}{{\ttfamily
  arXiv:2211.08448 [quant-ph]}}.

\bibitem{Bao_2022}
N.~Bao, C.~Cao, and G.~Zhu, ``{Deconfinement and error thresholds in
  holography},'' \href{http://dx.doi.org/10.1103/PhysRevD.106.046009}{{\em
  Phys. Rev. D} {\bfseries 106} no.~4, (2022) 046009},
  \href{http://arxiv.org/abs/2202.04710}{{\ttfamily arXiv:2202.04710
  [hep-th]}}.

\bibitem{aharonov1999faulttolerant}
D.~Aharonov and M.~Ben-Or, ``Fault-tolerant quantum computation with constant
  error rate,'' \href{http://dx.doi.org/10.1137/S0097539799359385}{{\em SIAM J.
  Comput.} {\bfseries 38} no.~4, (2008) 1207--1282},
  \href{http://arxiv.org/abs/quant-ph/9906129}{{\ttfamily
  arXiv:quant-ph/9906129}}.

\bibitem{Knill_1998}
E.~Knill, R.~Laflamme, and W.~H. Zurek, ``{Resilient quantum computation: Error
  models and threshold},'' \href{http://dx.doi.org/10.1098/rspa.1998.0166}{{\em
  Proc. Roy. Soc. Lond. A} {\bfseries 454} (1998) 365--384},
  \href{http://arxiv.org/abs/quant-ph/9702058}{{\ttfamily
  arXiv:quant-ph/9702058}}.

\bibitem{Kitaev_2003}
A.~Y. Kitaev, ``{Fault tolerant quantum computation by anyons},''
  \href{http://dx.doi.org/10.1016/S0003-4916(02)00018-0}{{\em Annals Phys.}
  {\bfseries 303} (2003) 2--30},
  \href{http://arxiv.org/abs/quant-ph/9707021}{{\ttfamily
  arXiv:quant-ph/9707021}}.

\bibitem{mayers2002superselection}
D.~Mayers, ``Superselection rules in quantum cryptography,''
  \href{http://arxiv.org/abs/quant-ph/0212159}{{\ttfamily
  arXiv:quant-ph/0212159 [quant-ph]}}.

\bibitem{verstraete2003quantum}
F.~Verstraete and J.~I. Cirac, ``Quantum nonlocality in the presence of
  superselection rules and data hiding protocols,''
  \href{http://dx.doi.org/10.1103/PhysRevLett.91.010404}{{\em Phys. Rev. Lett.}
  {\bfseries 91} (2003) 010404},
  \href{http://arxiv.org/abs/quant-ph/0302039}{{\ttfamily
  arXiv:quant-ph/0302039 [quant-ph]}}.

\bibitem{Kitaev:2003zj}
A.~Kitaev, D.~Mayers, and J.~Preskill, ``{Superselection rules and quantum
  protocols},'' \href{http://dx.doi.org/10.1103/PhysRevA.69.052326}{{\em Phys.
  Rev. A} {\bfseries 69} (2004) 052326},
  \href{http://arxiv.org/abs/quant-ph/0310088}{{\ttfamily
  arXiv:quant-ph/0310088}}.

\bibitem{Aharonov:1967zza}
Y.~Aharonov and L.~Susskind, ``{Charge Superselection Rule},''
  \href{http://dx.doi.org/10.1103/PhysRev.155.1428}{{\em Phys. Rev.} {\bfseries
  155} (1967) 1428--1431}.

\bibitem{bartlett2003entanglement}
S.~D. Bartlett and H.~M. Wiseman, ``Entanglement constrained by superselection
  rules,'' \href{http://dx.doi.org/10.1103/PhysRevLett.91.097903}{{\em Phys.
  Rev. Lett.} {\bfseries 91} (2003) 097903},
  \href{http://arxiv.org/abs/quant-ph/0303140}{{\ttfamily
  arXiv:quant-ph/0303140 [quant-ph]}}.

\bibitem{knill1997theory}
E.~Knill and R.~Laflamme, ``{A Theory of quantum error correcting codes},''
  \href{http://dx.doi.org/10.1103/PhysRevLett.84.2525}{{\em Phys. Rev. Lett.}
  {\bfseries 84} (2000) 2525--2528},
  \href{http://arxiv.org/abs/quant-ph/9604034}{{\ttfamily
  arXiv:quant-ph/9604034}}.

\bibitem{Gell-Mann:1960mvl}
M.~Gell-Mann and M.~Levy, ``{The axial vector current in beta decay},''
  \href{http://dx.doi.org/10.1007/BF02859738}{{\em Nuovo Cim.} {\bfseries 16}
  (1960) 705}.

\bibitem{PhysRev.175.2195}
M.~Gell-Mann, R.~J. Oakes, and B.~Renner, ``Behavior of current divergences
  under
  ${\mathrm{su}}_{3}\ifmmode\times\else\texttimes\fi{}{\mathrm{su}}_{3}$,''
  \href{http://dx.doi.org/10.1103/PhysRev.175.2195}{{\em Phys. Rev.} {\bfseries
  175} (1968) 2195--2199}.

\bibitem{10.1093/ptep/ptaa104}
{\bfseries Particle Data Group} {\bfseries Collaboration}, P.~A. Zyla {
  et~al.}, ``{Review of Particle Physics},''
  \href{http://dx.doi.org/10.1093/ptep/ptaa104}{{\em PTEP} {\bfseries 2020}
  no.~8, (2020) 083C01}.

\bibitem{ecker1994pionnucleon}
G.~Ecker, ``{The Pion - nucleon interaction as an effective field theory},''
  {\em Chin. J. Phys.} {\bfseries 32} (1994) 1303--1316,
  \href{http://arxiv.org/abs/hep-ph/9407240}{{\ttfamily arXiv:hep-ph/9407240}}.

\bibitem{GASSER1984142}
J.~Gasser and H.~Leutwyler, ``{Chiral Perturbation Theory to One Loop},''
  \href{http://dx.doi.org/10.1016/0003-4916(84)90242-2}{{\em Annals Phys.}
  {\bfseries 158} (1984) 142}.

\bibitem{SCHWINGER1957407}
J.~S. Schwinger, ``{A Theory of the Fundamental Interactions},''
  \href{http://dx.doi.org/10.1016/0003-4916(57)90015-5}{{\em Annals Phys.}
  {\bfseries 2} (1957) 407--434}.

\bibitem{Faist_2020}
P.~Faist, S.~Nezami, V.~V. Albert, G.~Salton, F.~Pastawski, P.~Hayden, and
  J.~Preskill, ``{Continuous symmetries and approximate quantum error
  correction},'' \href{http://dx.doi.org/10.1103/PhysRevX.10.041018}{{\em Phys.
  Rev. X} {\bfseries 10} no.~4, (2020) 041018},
  \href{http://arxiv.org/abs/1902.07714}{{\ttfamily arXiv:1902.07714
  [quant-ph]}}.

\bibitem{PhysRevD.19.3682}
E.~H. Fradkin and S.~H. Shenker, ``{Phase Diagrams of Lattice Gauge Theories
  with Higgs Fields},'' \href{http://dx.doi.org/10.1103/PhysRevD.19.3682}{{\em
  Phys. Rev. D} {\bfseries 19} (1979) 3682--3697}.

\bibitem{PhysRevA.105.052423}
M.~Mathur and A.~Rathor, ``{SU(N) toric code and non-Abelian anyons},''
  \href{http://dx.doi.org/10.1103/PhysRevA.105.052423}{{\em Phys. Rev. A}
  {\bfseries 105} no.~5, (2022) 052423},
  \href{http://arxiv.org/abs/2110.13841}{{\ttfamily arXiv:2110.13841
  [quant-ph]}}.

\bibitem{PhysRevB.79.045316}
F.~A. Bais and J.~K. Slingerland, ``{Condensate induced transitions between
  topologically ordered phases},''
  \href{http://dx.doi.org/10.1103/PhysRevB.79.045316}{{\em Phys. Rev. B}
  {\bfseries 79} (2009) 045316},
  \href{http://arxiv.org/abs/0808.0627}{{\ttfamily arXiv:0808.0627
  [cond-mat.mes-hall]}}.

\bibitem{Harvey_2020}
J.~A. Harvey and G.~W. Moore, ``{Moonshine, superconformal symmetry, and
  quantum error correction},''
  \href{http://dx.doi.org/10.1007/JHEP05(2020)146}{{\em JHEP} {\bfseries 05}
  (2020) 146}, \href{http://arxiv.org/abs/2003.13700}{{\ttfamily
  arXiv:2003.13700 [hep-th]}}.

\bibitem{Tachikawa_2015}
Y.~Tachikawa, \href{http://dx.doi.org/10.1007/978-3-319-08822-8}{{\em {N=2
  supersymmetric dynamics for pedestrians}}}.
\newblock Springer International, Cham, Switzerland, 2013.
\newblock \href{http://arxiv.org/abs/1312.2684}{{\ttfamily arXiv:1312.2684
  [hep-th]}}.

\bibitem{wess1992supersymmetry}
J.~Wess and J.~Bagger, {\em {Supersymmetry and supergravity}}.
\newblock Princeton University Press, Princeton, NJ, USA, 1992.

\bibitem{Furuya_2022}
K.~Furuya, N.~Lashkari, and S.~Ouseph, ``{Real-space RG, error correction and
  Petz map},'' \href{http://dx.doi.org/10.1007/JHEP01(2022)170}{{\em JHEP}
  {\bfseries 01} (2022) 170}, \href{http://arxiv.org/abs/2012.14001}{{\ttfamily
  arXiv:2012.14001 [hep-th]}}.

\bibitem{Furuya1_2022}
K.~Furuya, N.~Lashkari, and M.~Moosa, ``{Renormalization group and approximate
  error correction},''
  \href{http://dx.doi.org/10.1103/PhysRevD.106.105007}{{\em Phys. Rev. D}
  {\bfseries 106} no.~10, (2022) 105007},
  \href{http://arxiv.org/abs/2112.05099}{{\ttfamily arXiv:2112.05099
  [hep-th]}}.

\bibitem{mindist}
A.~Vardy, \href{http://dx.doi.org/10.1145/258533.258559}{``Algorithmic
  complexity in coding theory and the minimum distance problem,''} in {\em
  Proceedings of the Twenty-Ninth Annual ACM Symposium on Theory of Computing},
  STOC '97, p.~92–109.
\newblock Association for Computing Machinery, New York, NY, USA, 1997.

\end{thebibliography}\endgroup

\appendix
\section{Transition amplitude}\label{pion}

The effective Lagrangian describing the interactions between pions and nucleons, derived from the spontaneous breaking of axial $SU(2)$ symmetry, can be expressed as:
\begin{equation}
L_{\text{int}} = \int \dee^3x ~ g_1\sum_i \bar{N}\tau^i\gamma^5\slashed{\partial}\pi^i N + g_2\sum_i \bar{N}\tau^i\gamma^5\pi^i N,
\end{equation}
where $N=\begin{pmatrix} p\\ n \end{pmatrix}$ represents the nucleon field.

The strength of the interaction between pions and nucleons is determined by the coupling constants $g_1$ and $g_2$. These constants are proportional to $\frac{1}{f_{\pi}^2}$, where $f_{\pi}\sim 100~\text{MeV}$ denotes the pion decay constant. Consequently, pions scatter at a rate proportional to $(\frac{E}{\Lambda_{\text{QCD}}})^2$. Errors arise from transitions between proton and neutron states, which occur when a pion is absorbed or emitted. A relevant Feynman diagram illustrating this process is presented below. 
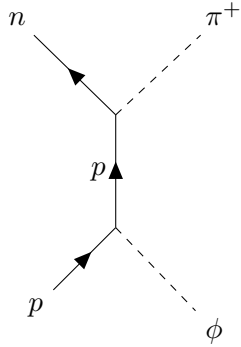
\begin{figure}[H]
\centering
\begin{tikzpicture}
\begin{feynman}
    \vertex (a) {\(p\)};
    \vertex [above right= of a](b);
    \vertex [above  =of b](c);
    \vertex [ below right= of b] (d){\(\phi\)};
    \vertex[above left=of c] (e) {\(n\)};
    \vertex[above right=of c] (f){\(\pi^+\)};
    \diagram*{
    (a)--[fermion](b)--[fermion,edge label=\(p\)](c),
    (c)--[fermion](e),
    (b)--[scalar](d),
    (c)--[scalar](f),
    };
\end{feynman}
\end{tikzpicture}
\caption{Flipping error induced by pion scattering.}
\end{figure}

The amplitude for the process can be estimated as follows:
\begin{equation}
\begin{split}
\mathcal{A}_{p\rightarrow n} = (-i\lambda_i)\bar{u}_n(k_3)(-ig_1\slashed{k}_4\gamma_5-g_2\gamma_5)\frac{i}{\slashed{k}_1+\slashed{k}_2-m_p}u_p(k_1)
\end{split}
\end{equation}
Here, $k_1$ through $k_4$ represent the momenta of the scattering particles. The masses of the particles are given by:
\begin{equation}
k_1^2=m_p^2, \quad k_2^2=m_{\phi}^2, \quad k_3^2=m_{n}^2, \quad k_4^2=m_{\pi}^2
\end{equation}

To account for the energy conservation constraint, we incorporate a delta function and perform integration over the outgoing momentum to calculate the total cross section:

\begin{equation}
\begin{split}
\sigma_{\text{tot}} &= \frac{1}{(2E_1)(2E_2)|v_1-v_2|}\int \frac{d^3k_3}{(2\pi)^3}\int \frac{d^3k_4}{(2\pi)^3} \frac{|\mathcal{A}_{p\rightarrow n}|^2}{(2E_3)(2E_4)}(2\pi)^4\delta^{(4)}(k_1+k_2-k_3-k_4)\\
&= \frac{1}{64\pi^2E_1E_2|v_1-v_2|} \int \dee\Omega |\mathcal{A}_{p\rightarrow n}|^2 \int \dee|\vec{k}_3| \frac{|\vec{k}_3|^2}{E_3}\frac{1}{E_4}\delta(E_3+E_4-E_{\text{CM}})\\
&= \frac{1}{64\pi^2E_{\text{CM}}^2} \frac{|\vec{k}_3|}{|\vec{k}_1|}\theta(E_{\text{CM}}-m_3-m_4) \int \dee\Omega |\mathcal{A}_{p\rightarrow n}|^2
\end{split}
\end{equation}

This results in the inclusion of a Heaviside step function $\theta(E_{\text{CM}}-m_n-m_{\pi})$, where $E_{\text{CM}}$ is the center of mass energy.  For a stationary target proton, a non-zero scattering cross section requires the incident energy $E_2$ of the scalar particle $\phi$ to satisfy $E_2 > m_{\pi}+\frac{m_{\pi}^2}{2m_{p}}$.

\end{document}